\documentclass[aps,prb,reprint,superscriptaddress,showpacs]{revtex4-1}

\usepackage{amsmath,amssymb}
\usepackage{graphicx}
\usepackage{color}
\usepackage{epstopdf}
\definecolor{darkgreen}{RGB}{34, 139, 34}

\usepackage[normalem]{ulem}

\renewcommand{\d}{{\rm d}}

\newcommand{\imai}{{\rm i}}

\begin{document}

\title{Superlattice gain in positive differential conductivity region}
\author{David O. Winge}
\email[]{Electronic mail: David.Winge@teorfys.lu.se}
\author{Martin Francki\'e}
\author{Andreas Wacker}
\date{\today}
\affiliation{Mathematical Physics, Lund University, Box 118, 22100 Lund, Sweden}

\begin{abstract}
We analyze theoretically a superlattice structure proposed by A. Andronov 
et al. [JETP Lett 102, 207 (2015)] 
to give Terahertz gain for an operation point with positive differential
conductivity. Here we confirm the existence of gain and show that an
optimized structure displays gain above 20 cm$^{-1}$ at low temperatures, so
that lasing may be observable. Comparing a variety of simulations,
this gain is found to be strongly affected 
by elastic scattering. It is shown that the
dephasing modifies the nature of the relevant states, so that the 
common analysis based on Wannier-Stark states is not reliable for a 
quantitative description of the gain in structures with extremely 
diagonal transitions. 
\end{abstract}


 
\maketitle

\begin{figure}
\centering
\includegraphics[width=0.9\columnwidth]{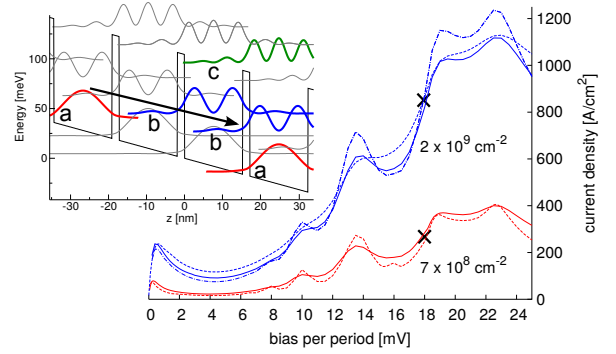}
\caption{Current-voltage characteristics for 
two SLs with different doping densities (full lines).
Simulations with modified scattering parameters are displayed as
dashed lines, and the one simulation at 40~K is shown as a dotted-dashed blue line. 
In the inset the configuration of the Wannier states at 18 mV is shown, 
marked by a cross in the main plot.
The gain transition studied is indicated by an arrow.}
\label{FigIV}
\end{figure}

Semiconductor superlattices \cite{EsakiIBM1970} (SLs) had always been considered as
an interesting candidate for THz gain materials due to the Bloch gain
\cite{KtitorovSovPhysSolState1972}, which was finally experimentally confirmed
more than 30 years later\cite{ShimadaPRL2003,SavvidisPRL2004,SekinePRL2005}. 
However, this type of gain
is intrinsically connected with the negative differential conductivity in the
current-field relation, so that the formation of field
domains\cite{EsakiPRL1974,GrahnBook1995,WackerPhysRep2002,BonillaRepProgPhys2005} strongly limits
its observation and practical use. As an alternative, it was suggested 
\cite{AndronocJPhysConfSer2009} that gain can be
present in the positive differential conductivity region of SLs
where resonant tunneling over several barriers
\cite{SchneiderPRL1990,Sibille_PRL1998,HelmPRL1999} is relevant.
The idea is to operate the SL slightly 
below the tunneling resonance from the ground state of well $\mu$ 
to the excited state in the next-neighboring well $\mu+2$ (see the
inset of FIG.~\ref{FigIV}), which guarantees positive differential 
conductivity. At the same time, gain is suggested for the strongly diagonal
transition to 
the excited  state in the well $\mu+3$, which is actually lower
in energy than the ground state in well $\mu$.
More detailed experimental studies confirmed the suggested shape of the
current-field relation, but were not conclusive with respect to THz
gain\cite{AndronovJETPLett+2015}. Thus the question remains, whether this type
of gain exists at all and whether it is strong enough to overcome losses. 
In order to address this question, we performed detailed simulations
with our non-equilibrium Green's function (NEGF) simulation scheme
\cite{WackerQuantEl2013}, which are reported here.
We find that this particular gain mechanism exists, but that it 
is not particularly strong for the structure proposed. 
Testing different doping densities and layer sequences, we observe gain above
20/cm at low temperatures, which could overcome losses in typical THz
waveguides \cite{FaistBook2013}.
We noticed that dephasing strongly reduces this type of
gain with an extremely diagonal transition. This can be quantified by the 
eigenstates of the lesser Green's function, which represent better states to
estimate gain than the conventional eigenstates of the Hamiltonian called 
Wannier-Stark (WS) states.

The NEGF model
allows for a self-consistent evaluation of the transport
with respect to both elastic and inelastic scattering as well as 
interactions with an electromagnetic field in semiconductor heterostructure
devices \cite{LeePRB2006,SchmielauAPL2009,KubisPRB2009,
HaldasIEEE2011,GrangePRB2015}. In particular, it
is suitable for the study of semiconductor SLs, as
it contains simpler approaches, such as  miniband transport
\cite{EsakiIBM1970}, Wannier-Stark hopping \cite{TsuPRB1975,CaleckiJP1984}, or
sequential tunneling \cite{KazarinovSPS1971} as limiting cases
\cite{WackerPRL1998}. 

In NEGF models, scattering is treated by self-energies that are evaluated 
self-consistently until convergence is reached. These objects are
functions of both momentum and energy, but in our implementation
 they are effectively treated as only energy dependent, and evaluated
at a representative set of momentum transfers for the scattering matrix 
elements\cite{WackerQuantEl2013}.
This set is chosen by a typical energy transfer 
$E_\mathrm{typ}=3 \, \textrm{meV} + 0.5 k_BT$, 
fitted to give scattering matrix elements matching those
calculated with thermalized subbands for other low doped
heterostructures. Here, we apply also different 
values, in order to mimic increased or 
decreased scattering environments. 
In this study all samples considered were assumed to be 
homogeneously doped. Unless stated otherwise, we also keep the
lattice temperature fixed at $T=77$ K, where we consider the
model to be both robust and accurate.

\begin{figure}
\centering
\includegraphics[width=0.8\columnwidth]{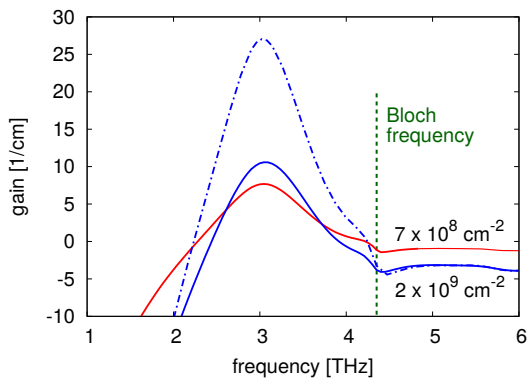}
\caption{Simulated gain for the two doping densities studied, both at 
a bias per period of 18 mV. For the higher doping density, gain at 40 K
lattice temperature is also shown (dotted-dashed). 
There are small signatures of dispersive gain around the Bloch frequency 
and the structure is mainly transparent at higher frequencies. }
\label{FigGain}
\end{figure}

FIG.~\ref{FigIV} shows the calculated current-voltage characteristics for the
device of Ref.~\onlinecite{AndronovJETPLett+2015} (red solid line for a doping of $7\times
10^8/\mathrm{cm}^2$ per period). The peak structure agrees reasonably well
with the experimental data shown in FIG.~4 of
Ref.~\onlinecite{AndronovJETPLett+2015}. For comparison, the experimental shoulder
at 19 mV per period, where the ground state is in resonance with the excited 
state of the 2nd nearest neighbor well, shows a current density of 
$450 \, \mathrm{A/cm}^2$. In the following we focus on the operation
point at 18 mV per period, which is a stable operation point with 
positive differential conductivity. The inset in FIG.~\ref{FigIV} shows the
Wannier levels at this field. 
FIG.~\ref{FigGain} shows the calculated gain (at weak cavity field). For the
nominal structure (red solid line) it remains well below 10/cm, which is probably 
too small to overcome the total losses. 

\begin{figure}
\centering
\includegraphics[width=1.0\columnwidth]{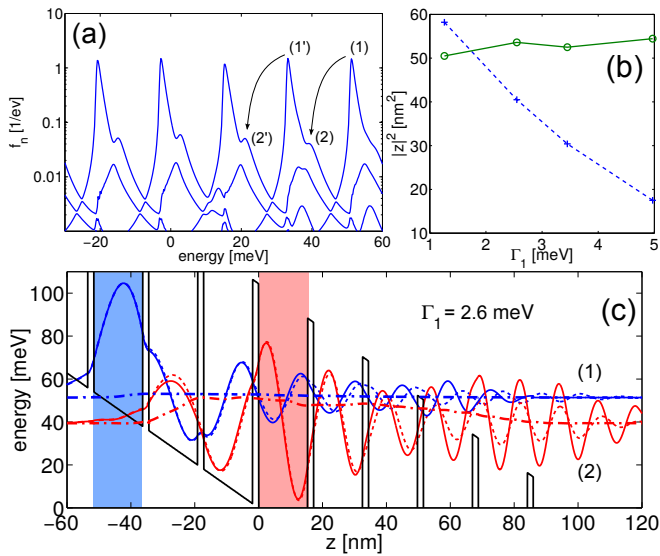}
\caption{(a) Eigenvalues $f_n(E)$ of the lesser Green's function 
$(2\pi\imai) \, G^<_{\alpha\beta}({\bf k}=0,E)$ at each energy point. 
(b) Modulus square of the dipole matrix elements against the energy broadening of the ground state. The eigenstates
of the density matrix (dashed blue) is strongly dependent on scattering as opposed to WS states (solid green). (c) Real part of the eigenstates (dashed) corresponding to the eigenvalues 
indicated in (a). The imaginary part is visualized 
by plotting the current carrying combination\cite{LeePRB2006} 
$\Re \{ -\imai \phi^* \d \phi /\d z /m^* \} $  
(dotted-dashed) for both eigenstates. These can be seen, 
especially for state (2), to extend over several periods.
For easy comparison we plot also the WS states 
$(\mu,i)$ (blue solid line) and  $(\mu+3,ii)$ (red solid line). 
The well where each wavefunction has its origin is
shaded as guidance.  }
\label{FigSuper}
\end{figure}

In the following we will employ a strict naming convention for the
SL states $(\mu,\nu)$ where $\mu$ will give the period
index, with 0 for the central period, and $\nu$ for the state index. 
Here, Wannier states, are denotes by letters $\nu=a,b,c$ and
Wannier-Stark states (WS states, which are the eigenstates of the Hamiltonian) 
by roman numbers $\nu=i,ii,iii$.
The tunneling resonance at the current peak at 19 mV per period is thus between
Wannier levels $(\mu,a)$ and $(\mu+2,b)$. 
At 18 mV per period the resonance between these levels is slightly detuned, 
so that the WS state $(\mu,i)$ is dominated by $(\mu,a)$ but has
significant admixtures from $(\mu+2,b)$ and $(\mu+1,b)$.
Similarly, the WS state $(\mu+2,ii)$ is dominated by $(\mu+2,b)$ with
significant admixtures from $(\mu,a)$, $(\mu+1,b)$, and $(\mu+3,b)$. 
These states are displayed in FIG.~\ref{FigSuper}~(c) by full lines.
The state $(\mu,i)$ is lower in energy than $(\mu+2,ii)$ and has 
thus a significantly larger occupation. 

Now the state $(\mu+3,ii)$, which is equivalent to $(\mu+2,ii)$, but shifted
to the right and down in energy, is about 14.7 meV below the state $(\mu,i)$.
As both states extend over several periods they overlap significantly
and furthermore there is inversion for the corresponding transition.
We can attribute the gain shown in FIG.~\ref{FigGain} to this transition,
where a slight red shift can be explained by 
dispersive gain \cite{TerazziNaturePhys2007}. 

As an attempt to improve inversion and gain, the doping was increased
to give a sheet density three times higher than the nominal 
sample. The result on current and gain is shown in FIG.~\ref{FigIV} 
and FIG.~\ref{FigGain}, respectively. 
As expected the current density increases approximately by a factor three. 
However, the  peak gain increases only slightly at 77 K. Significantly higher
values are found at lower temperatures, where our model suggests gain above
20/cm at 40 K.\footnote{However, we refrain from making a definite statement
  on specific values, as our model showed inaccuracies for some quantum
  cascade lasers at such low temperatures.}
Furthermore, in both samples there are small
signatures of Bloch gain at around 4.2 THz and we also observe that the 
high doped sample has more dark absorption at frequencies
far from the gain transition.

In the following, we want to study, why the increase of gain with doping is limited, 
so that its practical use appears questionable. A naive guess, would be
an increase of gain by a factor three just like the current. However,
the inversion might not be proportional to the doping and the linewidth
changes with doping. In order to study these effects, we use the standard
estimate for the gain using Fermi's Golden Rule (FGR) 
\begin{align}
G(\omega) = \frac{\Delta E_{fi}}{\hbar} 
\frac{e^2\Delta n_{fi} z_{fi}^2}{2n_r c \epsilon_0 d} 
\frac{\Gamma_w}{(\Delta E_{fi}-\hbar \omega)^2 + \Gamma_w^2/4}
\label{EqGFGR}
\end{align}
where $\Delta E_{fi}$ is the energy difference between the initial 
and final states, $\Delta n_{fi}$ is the inversion, 
$z_{fi}$ the dipole matrix element, $n_r$ is the refractive index and 
$\Gamma_w$ is the full width half maximum of the gain peak. 
These variables can be extracted from the full NEGF model where
we diagonalize the Hamiltonian including the real parts of the self-energies,
on the diagonal in order to shift the single particle energy levels, 
to get the WS states. 
Here we
approximate the linewidth as the sum of the lifetimes of the two 
states involved, $\Gamma_w=\gamma_f+\gamma_i$.  

\begin{table}
\centering
\setlength{\tabcolsep}{6pt}
\begin{tabular}{l | c c | c c | c}
\hline \hline
$\rho$ [1/cm$^2$] & \multicolumn{2}{c|}{$7\times 10^8$} & \multicolumn{2}{|c}{$2\times 10^9$} & \multicolumn{1}{|c}{$2\times 10^9$}  \\
\hline
$ E_{\rm typ}$ (meV) & 14 & 6.2 & 6.2 & 3.2 & 4.7 (40 K) \\
$\Gamma_1$(meV) & 1.3 & 2.6 & 3.4 & 5.0 & 2.3 (40 K) \\
\hline
NEGF &  19.2 & 7.30 & 9.50 & -1.64 & 25.9 \\
FGR(WS) & 20.7 & 11.8 & 26.6 & 19.8 & 37.5 \\
FGR($G^<$) & 23.8  & 8.90  & 15.4  & 6.39 & 29.2 \\
\hline \hline
\end{tabular}
\caption{Estimated gain in units cm$^{-1}$ from the gain transition using FGR with WS states and states from diagonalization of the lesser Green's function $G^<$, 
compared to the full NEGF calculations. 
In addition the lifetime broadening of the ground state $\Gamma_1$ is shown as it has 
a direct relation to the dephasing strength, as well as the $E_{\rm typ}$ parameter used for each simulation. }
\label{TabGain}
\end{table}

The result of this estimate is shown in TAB.~\ref{TabGain} for a set of
different model systems. The second and third column of TAB.~\ref{TabGain}
refer to our standard simulation parameters with $E_{\rm typ}=6.2$ meV at 77~K, as used in
FIGS.~\ref{FigIV}-\ref{FigGain} (full lines). Furthermore, we also performed simulations
with altered $E_\mathrm{typ}$. 
The data in the first/fourth column are for 
decreased/increased scattering compared to their
neighboring column. This is reflected by the respective width of the
ground state $\Gamma_1$, which is extracted from the NEGF
calculation. 
The current simulations for these parameters are shown by dashed lines in
FIG.~\ref{FigIV}. The minor changes in current can be understood by a slight
broadening/sharpening of the tunneling resonances for increased/decreased
scattering, respectively. The fifth column in TAB.~\ref{TabGain}  gives
results for  40 K  using our standard temperature dependent $E_{\rm typ}$.

Let us first consider the estimate from FGR
(\ref{EqGFGR}) with the common WS states in TAB.~\ref{TabGain}. Here we find, that the peak
gain follows essentially the doping density divided by $\Gamma_1$,
which shows that the inversion is essentially proportional to doping,
and all other ingredients, except for the broadening, are constant. In
contrast, the NEGF calculation shows a much stronger decrease of gain
with $\Gamma_1$. While a part of the differences may be attributed to
the widening of other absorbing transitions, the large extent is
stunning. 

To understand this discrepancy we analyze the eigenstates of the
lesser Green's function $G^<_{\alpha\beta}({\bf k},E)$ (which can be viewed as
the energetically resolved density matrix) and the corresponding wavefunctions
following Ref.~\onlinecite{LeePRB2006}.
The eigenvalues for the nominal case
with $\Gamma_1=2.6$ are plotted
in FIG.~\ref{FigSuper}~(a) for ${\bf k}=0$. 
The eigenvalues show two sets of peaks, $(\mu,1)$ and $(\mu,2)$, corresponding
to the ground and excited level. They also visualize the inversion at an
energy of 12 meV (indicated by arrows), corresponding to 3 THz.
As the eigenvalues are sorted by size in the diagonalization process,
we see anti-crossings where the different eigenstates passes each other,
so that the state at the eigenvalue indicated by (2) is not the same as
the one at (1) since they are separated by at least one anti-crossing.

\begin{figure}
\centering
\includegraphics[width=1.0\columnwidth]{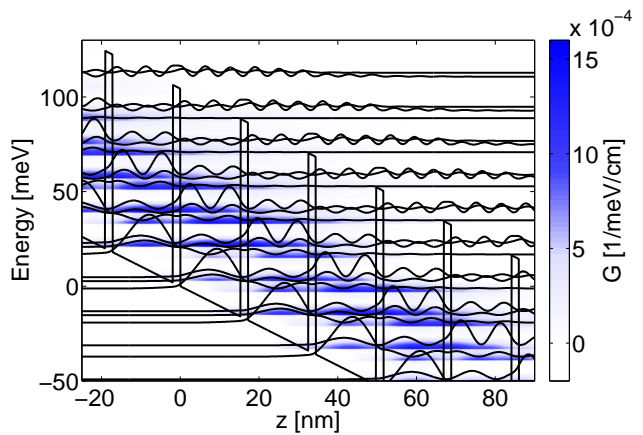}
\caption{Gain at 3 THz resolved in energy and space for the 
nominal case with $\Gamma_1=2.6$ meV.
The coherence giving rise to the gain extends over
several periods. The bias is 18 mV per period.}
\label{FigResolve}
\end{figure}
At the eigenvalue peaks (1) and (2) we plot the corresponding
eigenstates in FIG.~\ref{FigSuper}~(c) together with the WS states. From
this plot it is possible to see that compared to the WS states, 
the wavefunction corresponding to the eigenstate $(\mu+3,2)$ is 
slightly more localized than the WS state $(\mu+3,ii)$.
For these simulation parameters this leads to a decrease 
of the dipole matrix element. 
In FIG.~\ref{FigSuper}~(b) the modulus
square of the dipole matrix elements  are plotted versus the width $\Gamma_1$ of
the ground state. The WS states show small variations due to meanfield
and renormalization due to scattering, but are otherwise constant. In
contrast, the dipole matrix elements calculated by the eigenstates
of the Green's function, are comparable at low scattering but
provide a strong decrease with increasing
scattering. 
These dipole matrix elements can be applied in
FGR~(\ref{EqGFGR}), and the results are given in the lowest line of
TAB.~\ref{TabGain}. They actually follow the trend of the
full calculation, which demonstrates the relevance of these
eigenstates. The result by FGR naturally overestimates the gain
slightly, as we consider the gain from only one transition while
all other (mostly absorbing) transitions are
fully taken into account in the NEGF model.

The strong $\Gamma$-dependence of the eigenstates
of the lesser Green's function is reflected by dephasing, 
which affects the coherence length. In this particular situation, 
the gain is highly diagonal and is thus dependent on these 
spatial coherences. This is further demonstrated in 
FIG.~\ref{FigResolve}, where the gain stripes extend over more than 50 nm.

As a complement to changing doping, we also tried to optimized the SL
by modifying the well and barrier widths. Here we present results for 
the structures \textit{wide}/\textit{narrow}, where the well is
increased/decreased
by 4 monolayers, respectively, and the structure 
\textit{thin}, where the barrier is decreased by 2 monolayers compared to the nominal sample.
The sheet doping density was kept constant at $7\times 10^8/\mathrm{cm}^2$. 
\begin{figure}
\centering
\includegraphics[width=0.9\columnwidth]{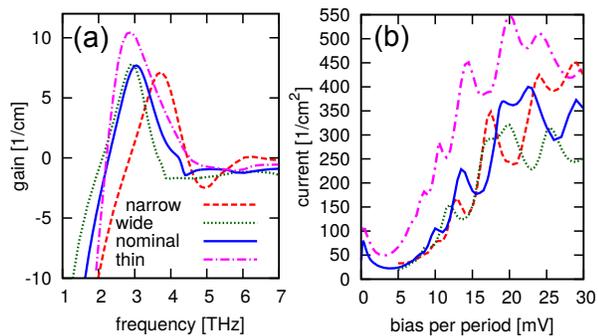}
\caption{(a) Gain and current (b) for standard scattering parameters for the three 
altered structures as well as the nominal sample from Ref.~\onlinecite{AndronovJETPLett+2015}.
The operating bias per period was 16 mV and 23 mV for the \textit{wide} and \textit{narrow} structure,
respectively. For the \textit{thin} structure the bias was fixed at 18.5 mV in the gain simulation. }
\label{FigAddMin}
\end{figure}
In FIG.~\ref{FigAddMin} we display gain and current for the samples,
which shows that adjusting the well width merely causes a shift in the peak frequency,
while thinner barriers improve the performance of the gain medium slightly.  


In {\em conclusion}, we have shown that the NEGF model predicts gain in the structure from 
Ref.~\onlinecite{AndronovJETPLett+2015}. However, the value is below 10/cm at
77~K, which hardly allows for lasing due to waveguide losses. Increasing the
doping, lasing at 40~K appears feasible. Further slight optimization by 
reducing the thickness of the barriers may be possible.

For this highly diagonal transition, the gain is strongly dependent on the 
scattering. This can be demonstrated by the
eigenstates of the lesser Green's function, which essentially differ
from the WS states in this case. We demonstrated that these unconventional 
states are more appropriate to calculate the dipole matrix elements 
for a quantitative description of gain by Fermi's golden rule.

{\em Acknowledgments:} We thank A. Andronov and J. Faist for helpful discussions and the
Swedish Research Council for financial support.

%


\end{document}